# Behavioural – based modelling and analysis of Navigation Patterns across Information Networks

Vesna Kumbaroska[1] and Pece Mitrevski[2]

[1] University of information science and technology, St. Paul the Apostle, Partizanska bb,
6000 Ohrid, Macedonia
vesna.gega@uist.edu.mk

[2] University St. Clement of Ohrid, Faculty of information and communication technologies,
Partizanska bb,
7000 Bitola, Macedonia
pece.mitrevski@fikt.uklo.edu.mk

**Abstract.** Navigation behaviour can be considered as one of the most crucial aspects of user behaviour in an electronic commerce environment, which is very good indicator of user's interests either in the process of browsing or purchasing. Revealing user navigation patterns is very helpful in finding out a way for increasing sale, turning the most browsers into buyers, keeping costumer's attention, loyalty, adjusting and improving the interface in order to boost the user experience and interaction with the system. In this regard, this research has identified the most common user navigation patterns across information networks, illustrated through the example of an electronic bookstore. A behavioural-based model that provides profound knowledge about the processes of navigation is proposed, specifically examined for different types of users, automatically identified and clustered into two clusters according to their navigational behaviour. The developed model is based on stochastic modelling using the concept of Generalized Stochastic Petri Nets which complex solution relies on Continuous Time Markov Chain. As a result, calculation of several performance measures is performed, such as: expected time spent in a transient tangible marking, cumulative sojourn time spent in a transient tangible marking, total number of visits in a transient tangible marking etc.

## 1. Introduction

The growing popularity of electronic commerce leads to an increased number of buyers and very frequent online purchasing. Every day, new online shopping standards are established, based on changed and grown user demands and expectations which implies a need to retrieve more precise results in the process of searching, accurate recommendations, better design oriented to the user, and adequate personalization. This requires a profound understanding of the interaction, i.e., the interface, on one hand, and the user behaviour, on the other hand, which is extremely important to create new strategies.

Navigation behaviour can be considered as one of the most important aspects of user behaviour in an electronic commerce environment, which also is very good indicator of user's interests. In this direction, the main idea of our work is to identify the most frequent navigation patterns and show how Petri Nets (PN), as transition based models, can be applied in modelling user navigation behaviour. In this work, we want to outline some intrinsic details of Generalized Stochastic Petri Nets (GSPN) application for describing user navigation patterns, which actually take some time to execute (perform).

The rest of the paper is organized as follows. In Section 2 we present related work in this field. Section 3 represents an overview of the Petri Net formalism and our contribution. A case study is presented in Section 4. In the last section, we give some conclusions and steps for future work.

## 2. Related Work

To build an effective user navigation behaviour model means to develop an accurate predictive mathematical model of the user behaviour. Usually, the models are based on log data, collected in a period of time [16, 20], which structure is complex, but it carries an important source of information about user behaviour. In order to study log data and to gain knowledge of how users navigate, statistical analysis and application of data mining techniques need to be performed. Usually, the emphasis is placed on developing models for: discovering user search or navigation patterns [24], predicting and proposing future user actions [7] and personalization based on user behaviour [2, 12, 21].

In order to discover common user navigation behaviour patterns and also common sequences of transitions, a data mining approach is used by [19], based on a particular case of log data, taking into account the duration of the website visits. Also, a data mining approach, but in combination with semi-Markov process in discrete time, is applied in [10], in order to understand and describe the user behaviour. They propose an algorithm for obtaining a transition probability matrix. The developed model is used to improve the design of the website and also for performance evaluation. In the study of [18] a new approach for predicting user behaviour in order to improve website performance is proposed, based on both log data and website structure. They apply Petri Nets in order to reveal the structure of the website and to predict next user action (path completition). Model based on Colored Petri Nets (CPN) [11] for predicting next user action is proposed by [12]. The approach is based on former user profiles and the current user session. Another application of Petri Nets formalism is found in the research of [22], but this time the emphasis is placed on modelling and analyzing the structure of the web site, so web pages are presented as states and transitions as arcs. Solving this model helps in predicting subsequent user actions in the navigation process (path complete). This research evolved, so in [17] the concept of Stochastic Time Petri Nets (STPN) [4] is implemented for modelling the website structure and predicting future user behaviour. Application of Stochastic Petri Nets [1] for modelling electronic store costumer behaviour in order to improve the quality of web services, their reliability, performance and availability is suggested by [14]. Detailed user online search and navigation behaviour in large enclosed spaces (ie. Shopping malls), is investigated by [25] in order to improve customer satisfaction when using the Internet and doing online shopping.

There are numerous studies related to the user navigation behavior clustering process. For example, [23] illustrates a new approach for clustering patterns of interests, based on registered navigational data from Chinese electronic store. Specifically, except the navigational paths the web page visit frequency and retention on a web page or a category are taken into consideration. Also, some researches in this area are directed to the construction and implementation of algorithms for sequences clustering where not only transitions, but their order is important and carries useful information about the user behaviour. One such approach is shown in the survey of [13]. They present a combination of standard clustering methods and techniques to analyze sequences based on Markov Chain, which is used to group user behaviour according to the similarity of the user actions order.

Although numerous researches related to different navigation behavior aspects exist, exploring the concept of Generalized Stochastic Petri Nets for modelling and analysis of navigation patterns across information networks has not been investigated to the best of found knowledge. Nevertheless, the stochastic nature of the navigation process is sufficient motive this research to be directed toward establishing a stochastic model that will fully correspond to the real picture in order to better understanding and visualization of user navigation behavior. Further, the navigation behavior could be considered as transition between states, which indicates that is quite natural to present the navigation process using the concept of Generalized Stochastic Petri Nets which complex solution relies on Continuous Time Markov Chain [8, 9].

## 3. Using Petri Nets to capture navigation behaviour patterns

The basic elements of PN are: places, transitions, tokens and arcs. Usually, places are presented as circles, transitions as rectangular boxes and tokens as black dots. Places are related to states, transitions are related to actions that can change the states and arcs determine directed relation between places and transitions [16]. The marking of PN is closely related to tokens and it is used to describe the dynamic behaviour of the system. In that context, a transition is enabled if all its input places contain at least one token. An enabled transition can fire by removing one (more) token from all its input places, and adding one (more) token in all its output places, following the arcs.

The behaviour in a standard PN is discrete only. It means all the transitions are instantaneous or fire instantly. An extension of this concept is GSPN [1], where immediate and timed transitions are introduced. Here, the firing delays of timed transition are stochastic, usually exponentially distributed random variables. It means a timer is associated to each enabled timed transition, in such way that the timer value is sampled from (negative) exponential distribution with appropriate rate parameter. The timer constantly decreases, and when its value will become zero, the timed transition will fire. Usually, immediate transitions are presented as black rectangular boxes or bars, and timed transitions as white rectangular boxes. The immediate transition fire with priority over timed transitions. The markings could be tangible and vanishing. A tangible marking is a marking where only timed transitions are enabled. Contrary to this, a vanishing marking is a marking where immediate transitions are enabled (or combination of immediate and timed transitions are enabled).

### 3.1. Suggested model

The dynamic nature of PN, especially GSPN, indicates that they can be accommodated for modelling real user navigation behaviour. The focus in our research is placed on discovering and modelling navigation behaviour of bookstore users, specifically examined in the case of the first Macedonian electronic bookstore (www.kupikniga.mk), but easily generalized and applicable on other much known services, which structure is similar to the selected scenario.

The first step is to construct the web site topology, which means creating representation of the web site from the data collected, as it is shown in Fig. 1.

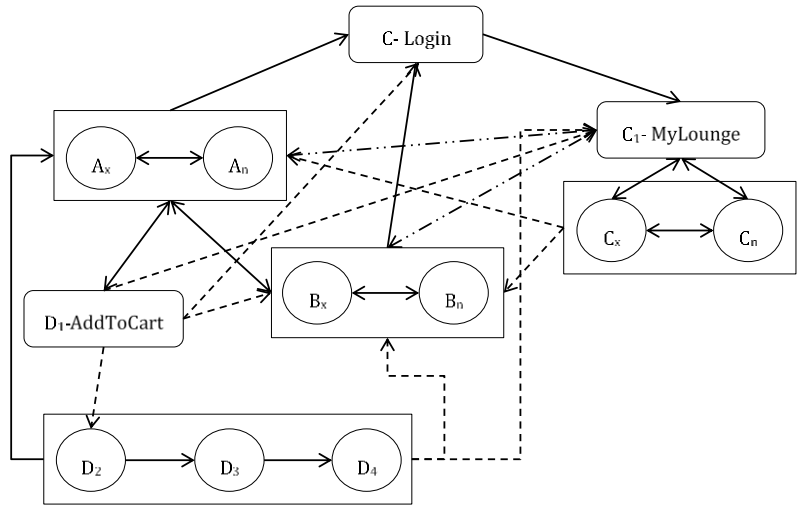

**Fig. 1.** Graphical model presentation

All web pages revealed are classified into four categories, based on their use. Detailed specification is given in Table 1. The 'A' category web pages are related to the products sold, in our case books. Web pages referred to general store information and help are categorized as 'B'. Next, the 'C' category web pages are only available for logged users, and they are associated with user's profile information. At the end, the 'D' category web pages are directly connected with the purchasing process.

**Table 1.** Detailed page specification

| Category | Page | Category | Page |
|---|---|---|---|
| A | Registration | C | LogIn |
|   | ListBooksforCategory |   | MyLoungeDef |
|   | Default |   | UserProfile |
|   | NewProducts |   | UserAddress |
|   | BestSellers |   | UserOrders |
|   | ProductDiscounts |   | SavedForLater |
|   | News |   | NewsLetter |
|   | Publisher |   | UserReferals |
|   | BookStoreMenu |   | ImportCSV |
|   | SearchPreview |   | HelpUser |
|   | ContactUs | D | AddToCart |
|   | Help |   | LastCartPreview |
|   | BookDetails |   | PayingCasys2 |
|   | ReadPDF |   | AddressEntry |
| B | AboutUs |   |   |
|   | GeneralProvisions |   |   |
|   | ProtectPersonalData |   |   |
|   | DeliveryPolitics |   |   |
|   | BackPolicy |   |   |
|   | HelpOrders |   |   |
|   | HelpRegistration |   |   |
|   | Marketing |   |   |

In GSPN notation, our model comprises 10 places and 44 timed transitions, as shown in Table 2. and Table 3., respectively.

**Table 2.** Places in the GSPN model

| Place name | Place description | Initial marking |
|---|---|---|
| A | A category page | 1 |
| B | B category page | 0 |
| E | Ended user session | 0 |
| L | Login | 0 |
| ML | MyLounge (MyBookStore) | 0 |
| C | C category page | 0 |
| D1 | AddToCart | 0 |
| D2 | AddressEntry | 0 |
| D3 | LastCartPreview | 0 |
| D4 | PayingCasys | 0 |

Table 3. Transitions in the GSPN model

| Transition name | Transition description | Rate |
| --- | --- | --- |
| tA_cont, tA1, tA2, tA3, tA4, tA5, tA6 | Visit an A category page | α |
| tB, tB_cont, tB1, tB2, tB3, tB4, tB5 | Visit a B category page | λ |
| tE_A, tE_B, tE_L, tE_ML, tE_C, tE_D1, tE_D2, tE_D3, tE_D4 | End the session | μ |
| tL, tL1, tL2, tL_cont | Login | κ |
| tML, tML1, tML2, tML3, tML4, tML5, tML_cont | Visit MyLounge page | ν |
| tC, tC_cont | Visit a C category page | θ |
| tD1, tD1_cont | Visit AddToCart page | ε |
| tD2, tD2_cont | Visit AddressEntry page | γ |
| tD3, tD3_cont | Visit LastCartPreview page | δ |
| tD4, tD4_cont | Visit PayingCasys page | β |

The GSPN dynamics of the model graphicaly depicted in Fig.2 can be described using the terminology of CTMC [1, 15].

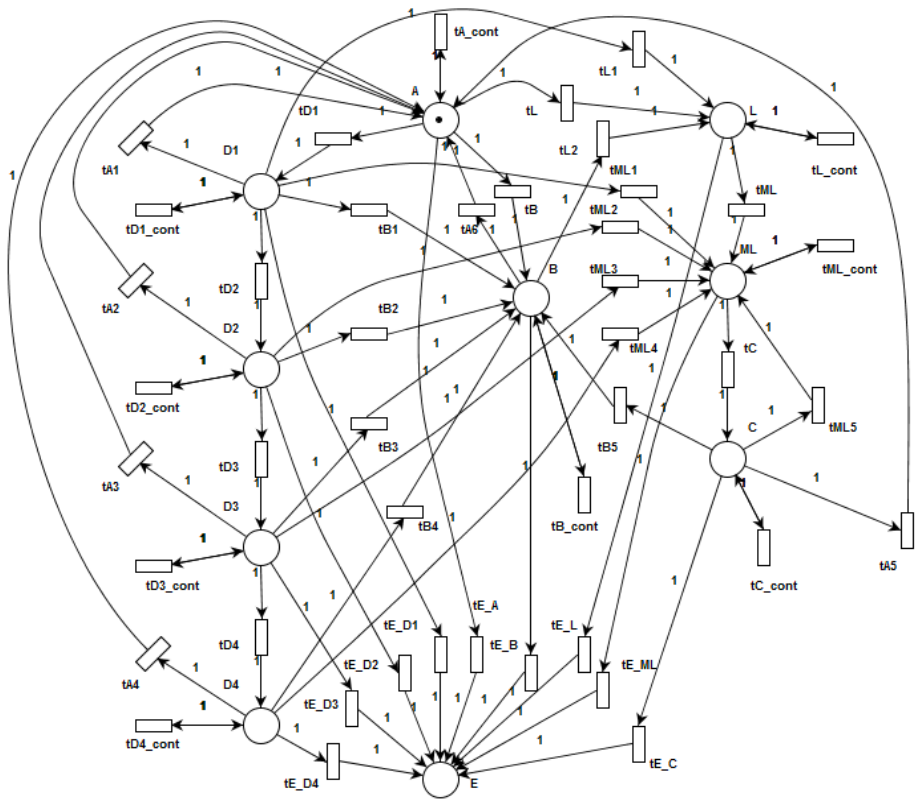

**Fig. 2.** Graphical model presentation

In this regard, for understanding the stochastic process that underlies this GSPN model and evaluation of appropriate performance measures, we use efficient time and space algorithm for computing steady state solutions of deterministic (DSPN) and stochastic (SPN) Petri Nets, proposed by [6].

## 4.    Case study

The navigation data used for this study is collected server side in a certain time frame and it is in a standard W3C format. The log file contains about 415000 records so that each record contains data belonging to several categories (attributes), but for this research particularly important categories are: userID, Date and Time and URL visited. In order to transform the data into easily interpretable format for further usage, it was necessary to do pre-processing and data cleaning tasks. It means removal of all incomplete records or records that lack any of the key attributes or they are inconsistent. We dynamically identify users, visits, and page views per user and per visit, as shown in Table 4.

**Table 4.** Information about the log file

| | | | |
|---|---|---|---|
| The total number of records analyzed | 322000 | | |
| Unique users | 1984 | | |
| Unique visits | 15433 | | |
| | **Mean** | **Min/Max** | **SD** |
| Time between two successive visits | 3d:20h:06m:59s | (0d:0h:0m:0s, 332d:23h:52m:57s) | |
| Visits per user | 7.78 | (1, 229) | 13.89 |
| Views per visit | 20.85 | (1, 2929) | 52.60 |

### 4.1.    Clustering methodology

After the user identification stage, the idea is to group the users based on their similar navigation behavior patterns that are considered as sequences of discrete events, in this case page views. In these sequences not only the user actions but also their order is important and carries useful information about the user and user's interests. For this purpose we implement the approach described in [13], where combination of standard clustering methods and techniques to analyze sequences based on Markov Chain is used. The first step in this approach is to create a model, which inputs are: a case table which is a standard table containing data related to the unique users. The second step is initialization of several parameters, most of them used with their default values, and one of them related to the number of resulting clusters set to 0, which means an optimal number of clusters according to the input data are generated. The third step is an evaluation of the model which means that each user with some probability is assigned to each of the generated clusters. The re-evaluation of the model continues until the algorithm converges. After this kind of set up, our users are automatically grouped into two clusters, as an optimal solution for our problem, which means two different groups of user navigation behavior are obtained. As illustrated in Fig.3, the number of users in each cluster is approximately equal.

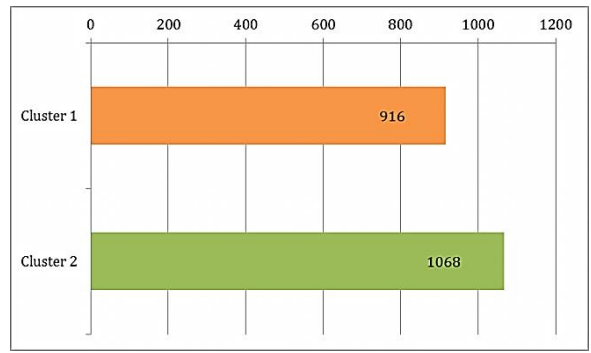

**Fig. 3.** Number of users in the two clusters

The visual cluster profiling is given in Fig.4, thus each column represents a cluster, each row represents a sequential attribute and each cell contains a histogram of user actions sequence.

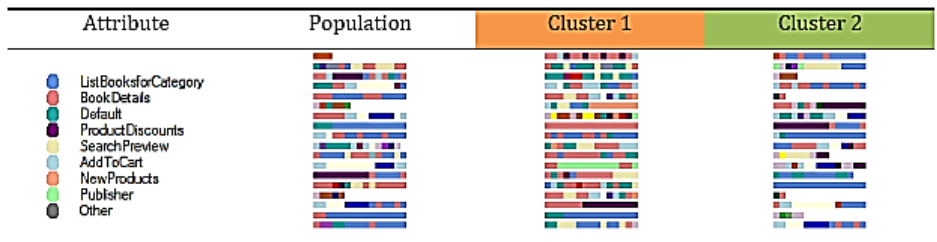

Fig. 4. Cluster profiles

From the visual inspection of Fig.4, we can notice that most prevalent action in the first cluster sequences is seeing book details as an A category page, versus the second cluster where users prefer filtering books by category, also as an A category page.

The most characteristic transitions and their probabilities are given in Fig.5 and Fig. 6, for the first and the second cluster respectively.

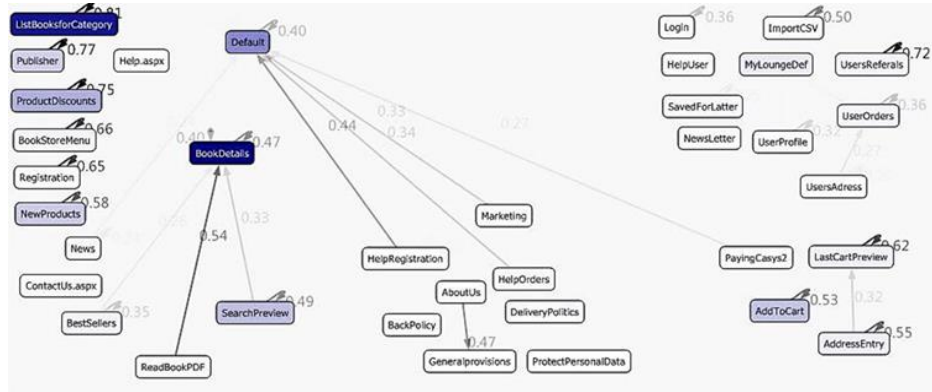

**Fig. 5.** Most characteristic transitions in Cluster 1

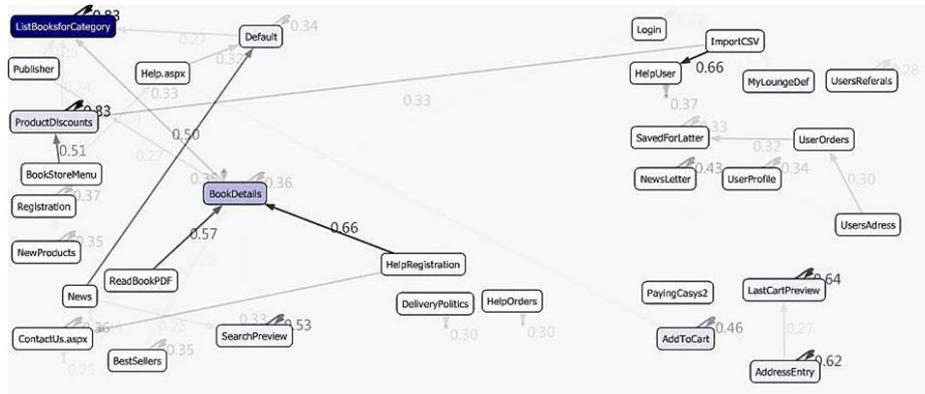

**Fig. 6.** Most characteristic transitions in Cluster 2

Actually, these diagrams represent Markov chains where each node is a sequential state and each arc is a transition from one state to another, which means it is directed. Also, each arc contains a weight which is related to the transition probability. The diverse color of the node is linked to the node popularity, which means more popular states are darker and vice versa. In this direction, we can discuss some of the most frequent navigation behavior patterns for the first cluster, demonstrated in Fig.5:

- 40% of the users start their visit seeing details for a book ('A' category). Around half of them (47%) choose to visit the same page again later.
- 22% of the users start their visit with the default page ('A' category). Also, around 40% of these users visit the same page again.
- Similar, around 20% of the users start their visit previewing their lounge ('C' category), and only 22% of them decide to review their orders.
- One third of the users that have bought a book ('D' category) choose to visit the default page ('A' category).
- Similar, 32% of the users that have changed the delivery address ('C' category), choose to see their cart ('D' category), but almost 62% of them decide to see their cart again later.

- etc.

Similar, some of the most frequent navigation behavior patterns for the second cluster, demonstrated in Fig.6 are:
- 35% of the users start their visit seeing details for a book ('A' category). One third of them (36%) choose to visit the same page again later.
- 22% of the users start their visit previewing their lounge ('C' category).
- 66% of the users which have invited friends ('C' category) choose to ask for help ('A' category) and 37% of those users end the visit.
- 27% of the users that have changed the delivery address ('C' category), choose to see their cart for the last time ('D' category), but almost 64% of this users decide to see their cart again later.
- etc.

## 4.2. Solving the GSPN model

Taking into consideration the both clusters characteristics, detailed analysis of the navigation patterns is performed. Based on the quantitative data obtained, the GSPN model is solved, validated and several performance measures are calculated. Firstly, the rates of our timed transitions are calculated and shown in Table 5.

Table 5. Firing rates of the timed transitions

|   | Cluster 1 | Cluster 2 |
|---|---|---|
| $\alpha$ | 0.000009 | 0.000033 |
| $\lambda$ | 0.000059 | 0.008621 |
| $\mu$ | 0.000003 | 0.000003 |
| $\kappa$ | 0.000004 | 0.000003 |
| $\nu$ | 0.000107 | 0.000080 |
| $\theta$ | 0.000882 | 0.001027 |
| $\varepsilon$ | 0.000073 | 0.000208 |
| $\gamma$ | 0.083333 | 0.050000 |
| $\beta$ | 0.047619 | 0.055556 |

Additionally, the corresponding state transition rate diagram of the CTMC is built and depicted in Fig.7.

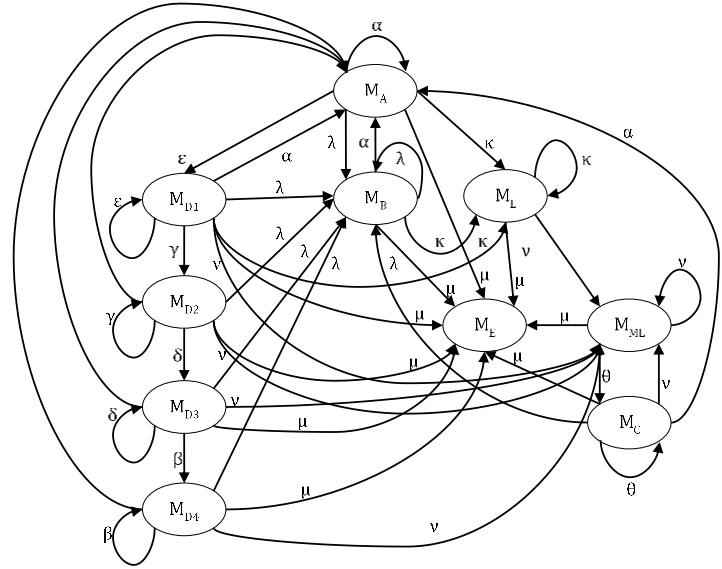

**Fig. 7.** State transition rate diagram of the CTMC

Because the last state (E) is absorbing it means that computing steady state probability distribution is meaningless in this case. All the others measures of interests are computed and described in the remainder of this section.

1) Average sojourn time

The average sojourn time in a given marking is a reciprocal value of the sum of the rates of the transitions enabled in that marking which lead the GSPN from that marking to other marking.

$$ST_i = \frac{1}{\sum_{t_j \in E(M_i)} rate_j} \qquad (1)$$

Here, $E(M_i)$ is related to the set of transitions enabled in marking $M_i$ and $rate_j$ corresponds to their exponentially distributed firing rates. The average sojourn times for our tangible states for both of the clusters are calculated and shown in Table 5.

**Table 6.** Average sojourn time spent in a transient tangible marking

|  | Cluster 1 | | Cluster 2 | |
| --- | --- | --- | --- | --- |
| Marking | Time [s] | Time [min] | Time [s] | Time [min] |
| $M_A$ | 6796.70 | 113.28 | 112.76 | 1.88 |
| $M_B$ | 13451.82 | 224.20 | 115.47 | 1.92 |
| $M_E$ | ∞ | ∞ | ∞ | ∞ |
| $M_L$ | 8820.74 | 147.01 | 11582.01 | 193.03 |
| $M_{ML}$ | 1008.46 | 16.81 | 901.15 | 15.02 |
| $M_C$ | 943.99 | 15.73 | 102.42 | 1.71 |
| $M_{D1}$ | 11.96 | 0.20 | 16.96 | 0.28 |
| $M_{D2}$ | 7.63 | 0.13 | 8.75 | 0.15 |
| $M_{D3}$ | 9.38 | 0.16 | 13.44 | 0.22 |
| $M_{D4}$ | 16.95 | 0.28 | 53.08 | 0.88 |

2) Expected time spent in a transient tangible marking

Expected time spent in a transient tangible marking before absorption is obtained as:

$$\vec{x}Q^N = -\vec{\pi}(0)^N \qquad (2)$$

Here, $Q^N$ and $\vec{\pi}(0)^N$ are restrictions of infinitesimal generator matrix $Q$ and the initial steady state probability vector $\vec{\pi}(0)$ respectively, to the set of transient markings $N$. The total time spent in the transient tangible states is calculated and given in Table 7.

**Table 7.** Total time spent in a transient tangible marking

|  | Cluster 1 | | Cluster 2 | |
| --- | --- | --- | --- | --- |
| Marking | Time [s] | Time [min] | Time [s] | Time [min] |
| $M_A$ | 32577.45 | 542.96 | 645.74 | 10.76 |
| $M_B$ | 338538 | 5642.3 | 142562.3 | 2376.04 |
| $M_L$ | 13496.14 | 224.94 | 23.42 | 0.39 |
| $M_{ML}$ | 8099.07 | 134.98 | 3.04 | 0.05 |
| $M_C$ | 40131.37 | 668.86 | 0.36 | 0.01 |
| $M_{D1}$ | 28.48 | 0.47 | 2.29 | 0.04 |
| $M_{D2}$ | 49.65 | 0.83 | 1.78 | 0.03 |
| $M_{D3}$ | 40.07 | 0.67 | 5.24 | 0.09 |
| $M_{D4}$ | 13241.97 | 220.7 | 6.06 | 0.1 |

3) Total number of visits in a transient tangible marking

Expected total number of visits in the transient states up to steady state can be computed from the following equation:

$$n_E(I - \Pi_E^*) = \pi_E^*(0) \qquad (3)$$

Here, $\Pi_E^*$ and $\pi_E^*$ are restrictions of the transition probability matrix $\Pi_E$ and the initial steady state probability vector $\vec{\pi}(0)$ respectively, to the set of transient markings $E$. The average number of visits in the transient tangible states is presented in Table 8.

**Table 8.** Average number of visits in a transient tangible marking

|        | Cluster 1 | Cluster 2 |
|--------|-----------|-----------|
| $M_A$    | 3.501551  | 11.593189 |
| $M_B$    | 3.828545  | 12.600198 |
| $M_L$    | 0.996941  | 1.062461  |
| $M_{ML}$ | 5.182163  | 1.035813  |
| $M_C$    | 5.164502  | 1.032780  |
| $M_{D1}$ | 1.839687  | 0.273193  |
| $M_{D2}$ | 1.835698  | 0.232543  |
| $M_{D3}$ | 1.828880  | 0.200941  |
| $M_{D4}$ | 1.823379  | 0.107744  |

4) Cumulative sojourn time in a transient tangible marking

Also, cumulative sojourn time is an interesting measure that could be calculated for the transient markings. In order to do that, for each marking the average number of visits should be found and it should be multiplied by the appropriate sojourn time, as follows:

$$\sigma_i = n_{Ei} * ST_i \qquad (4)$$

In our model, cumulative sojourn times for all transient markings are shown in Table 9.

**Table 9.** Cumulative sojourn time in a transient tangible marking

|         | Cluster 1 | | Cluster 2 | |
|---------|-----------|-----------|-----------|-----------|
| Marking | Time [s]  | Time [min] | Time [s] | Time [min] |
| $M_A$    | 23798.99 | 396.65 | 1307.2   | 21.79 |
| $M_B$    | 51500.88 | 858.35 | 1454.9   | 24.25 |
| $M_L$    | 8793.76  | 146.56 | 12305.43 | 205.09 |
| $M_{ML}$ | 5226.02  | 87.1   | 933.43   | 15.56 |
| $M_C$    | 4875.23  | 81.25  | 105.78   | 1.76 |
| $M_{D1}$ | 22.01    | 0.37   | 4.63     | 0.08 |
| $M_{D2}$ | 14       | 0.23   | 2.03     | 0.03 |
| $M_{D3}$ | 17.15    | 0.29   | 2.7      | 0.05 |
| $M_{D4}$ | 30.9     | 0.52   | 5.72     | 0.1 |

Thus, we can conclude that the average duration of a single session in the first cluster is 94278.95 [s] or 1571.32 [min], and 16121.82 [s] or 268.70 [min] in the second cluster.

## 5.    Conclusion

We have proposed a Generalized Stochastic Petri Nets (GSPN) modelling approach for describing user navigation behaviour. In this class of Petri Nets (PN), immediate transitions have firing weights, timed transitions have exponentially distributed firing times, and the underlying stochastic process is a Continuous Time Markov Chain (CTMC). In order to evaluate several performance measures, we have employed an efficient algorithm for computing steady state solutions of deterministic and stochastic Petri Nets (DSPN), proposed by [6]. The obtained results include the average sojourn time in each of the transient states, the total time spent in these transient states, the average number of visits, as well as the cumulative sojourn time.

Under the assumption that users' arrival can be described as a stochastic process with exponentially distributed interarrival times, and users' sessions have exponentially distributed duration with mean equal to the cumulative sojourn time, one can easily extend the GSPN model in order to describe the "operational environment", i.e. to partition the "input space" of the client-side by grouping users that exhibit as nearly as possible homogenous online navigation behaviour, as well as to evaluate service reliability, system availability, performance and performability. Moreover, during a session, all those multiple requests/tasks consume system resources, possibly reducing service accessibility (load levels saturate the system). An understanding of the motives underlying user actions can help designers to better accommodate to what appears to be chaos: make available those capabilities that best support the range of known behaviour patterns.

## 6.    Acknowledgment